\documentstyle[12pt]{article}

\makeatletter
   \newcounter{@sc}
   \newcounter{@scp}
   \newcounter{@t}
   \newlength{\@x}
   \newlength{\@xa}
   \newlength{\@xb}
   \newlength{\@y}
   \newlength{\@ya}
   \newlength{\@yb}
   \newsavebox{\@pt}

   \def\bezier#1(#2,#3)(#4,#5)(#6,#7){\c@@sc#1\relax
   \c@@scp\c@@sc \advance\c@@scp\@ne
   \@xb #4\unitlength \advance\@xb -#2\unitlength \multiply\@xb \tw@
   \@xa #6\unitlength \advance\@xa -#2\unitlength
   \advance\@xa -\@xb \divide\@xa\c@@sc
   \@yb #5\unitlength \advance\@yb -#3\unitlength \multiply\@yb \tw@
   \@ya #7\unitlength \advance\@ya -#3\unitlength
   \advance\@ya -\@yb \divide\@ya\c@@sc
   \setbox\@pt\hbox{\vrule height\@halfwidth depth\@halfwidth
   width\@wholewidth}\c@@t\z@
   \put(#2,#3){\@whilenum{\c@@t<\c@@scp}\do
   {\@x\c@@t\@xa \advance\@x\@xb \divide\@x\c@@sc \multiply\@x\c@@t
   \@y\c@@t\@ya \advance\@y\@yb \divide\@y\c@@sc \multiply\@y\c@@t
   \raise \@y \hbox to \z@{\hskip \@x\unhcopy\@pt\hss}\advance\c@@t\@ne}}}
 \makeatother

\begin{document} 
{ \pagestyle{empty} 

\vskip 30mm 

\centerline{\Large \bf The Structure of the Bazhanov-Baxter Model}
\centerline{\Large \bf and a New Solution of the Tetrahedron Equation}

\vskip 20mm


\vskip 20mm

\centerline{Minoru Horibe \footnote{E-mail :
horibe@edu00.f-edu.fukui-u.ac.jp}}
\centerline {{\it Department of Applied Physics, Faculty of Engineering}}
\centerline {{\it Fukui University, Fukui 910, Japan}}

\vskip 3mm

\centerline{Kazuyasu Shigemoto \footnote{E-mail :
shigemot@tezukayama-u.ac.jp}}
\centerline {{\it Department of Physics}}
\centerline {{\it Tezukayama University, Nara 631, Japan }}

\vskip 30mm

\centerline{\bf Abstract} 

We clarify the structure of the Bazhanov-Baxter model of the
3-dim $N$-state integrable model.
There are two essential points, i) the cubic symmetries, and
ii) the spherical trigonometry parametrization,
to understand the structure of this model.
We propose two approaches to find a candidate as a solution of the
tetrahedron equation, and we find a new solution.

\newpage
}
\section{Introduction}

Though there are many solutions of 2-dim integrable
statistical models, there are only a few solutions of 3-dim
integrable statistical models.
The first non-trivial example of a 3-dim integrable model was given
by Zamolodchikov. \cite{Zamolodchikov} This Zamolodchikov model is the
two colors string scattering model, and Baxter \cite{Baxter} reformulated
the
Zamolodchikov model into a $2$-state interaction around the cube
model and completed the proof that the Zamolodchikov model
satisfies the tetrahedron equation.

After these pioneering works, there was little progress for some time until
the
work of Bazhanov and Baxter. Bazhanov and Baxter \cite{B-B}
gave the integrable $N$-state interaction around the cube model,
which is the $N$-state generalization of the Zamolodchikov model.
This Bazhanov-Baxter model is constructed from the two principles
of i) interpreting the 2-dim $sl(n)$-generalized  chiral Potts
model \cite{B-K-M-S,D-J-M-M}
as a projected 3-dim model, and ii) comparing the $sl(n)$-generalized
 chiral Potts model with the Zamolodchikov model.
Bazhanov and Baxter have shown that two transfer matrices commute in their
model.

Later Kashaev et al. showed in a series of papers
that the Baxter-Bazhanov solution really satisfies the tetrahedron
 equation. \cite{Kashaev1,Kashaev2} We have checked
their proof in detail. \cite{ours1}

In addition to these works, there are many interesting papers
on the solutions of the tetrahedron
equation. \cite{Mangazeev1,Mangazeev2,Hietarinta,Korepanov,ours2}

Despite these works, only a few solutions have been found to this time.
In this situation, it will be necessary to find as many solutions as
possible before we try to understand the mathematical structure
behind the 3-dim integrable model.

In this paper, we first clarify the structure of the Bazhanov-Baxter model
and proposed two approaches to find a candidate as a
solution of the tetrahedron equation, and we find a new solution.

\section{The structure of the Bazhanov-Baxter model}

We first review the formulation of the Bazhanov-Baxter model
and give the condition of Kashaev et al., \cite{Kashaev2}
which is a sufficient condition to satisfy the tetrahedron
equation. Then we give an explicit form of the
solution, \cite{B-B,Kashaev1}
which is parametrized by using the angles of  spherical triangles.
Next, we clarify the cubic symmetries of the Bazhanov-Baxter
model in such a way that it gives a guiding principle to find
a candidate as a solution of the tetrahedron equation.
In addition to the cubic symmetries, the spherical trigonometry
parametrization is the key to understand the
structure of the Bazhanov-Baxter model.
We give two approaches, i) one using cubic symmetries
, and ii) an intuitive approach,
to the spherical trigonometry parametrization.

\subsection{Formulation of the Bazhanov-Baxter model}

Let us consider a simple cubic lattice and attach spin variables
to the lattice points. These spin variables
take the values $1,2,\cdots , N$, which we refer to an $N$-state interaction
around the
cube model. The Boltzmann weight with the spin variables
 $a$ - $h$ on the cube is given by
$W(a|e,f,g|b,c,d|h)$.

\unitlength 3pt
\begin{picture}
(50,50)
\put(51,11){$e$}           \put(71,11){$d$}
\put(39,20){$c$}           \put(60,20){$h$}
\put(51,31){$a$}           \put(71,31){$f$}
\put(39,41){$g$}           \put(59,40){$b$}

\put (50,10){\circle*{2}}
\put (70,10){\circle*{2}}
\put (38,19){\circle*{2}}
\put (58,19){\circle*{2}}
\put (50,30){\circle*{2}}
\put (70,30){\circle*{2}}
\put (38,39){\circle*{2}}
\put (58,39){\circle*{2}}

\thicklines
\put(50,10){\line(0,1){20}}
\put(70,10){\line(0,1){20}}
\put(38,19){\line(0,1){20}}
\put(50,10){\line(1,0){20}}
\put(50,30){\line(1,0){20}}
\put(38,39){\line(1,0){20}}
\put(50,10){\line(-4,3){12}}
\put(50,30){\line(-4,3){12}}
\put(70,30){\line(-4,3){12}}
\thinlines
\put(58,19){\line(0,1){20}}
\put(38,19){\line(1,0){20}}
\put(70,10){\line(-4,3){12}}
\put(30,2){{\bf Fig.~1}~ Spin assignment for the cube}

\end{picture}

\vspace{5mm}

The partition function of this model is given by

\begin{eqnarray}
   Z=\sum_{\rm spins} \prod_{\rm cubes} W(a|e,f,g|b,c,d|h).
\label{2.1}
\end{eqnarray}

The tetrahedron equation, which is the integrability condition of the
3-dim  statistical model, is given in the form

\begin{eqnarray}
   \sum^{N}_{d=1}  &&
    W(a_4|c_2,c_1,c_3|b_1,b_3,b_2|d)
    W'(c_1|b_2,a_3,b_1|c_4,d,c_6|b_4)
                                                    \nonumber \\
   &&\times W''(b_1|d,c_4,c_3|a_2,b_3,b_4|c_5)
   W'''(d|b_2,b_4,b_3|c_5,c_2,c_6|a_1)
                                                     \nonumber \\
   =\sum^{N}_{d=1} &&  W'''(b_1|c_1,c_4,c_3|a_2,a_4,a_3|d)
   W''(c_1|b_2,a_3,a_4|d,c_2,c_6|a_1)             \nonumber \\
   &&\times  W'(a_4|c_2,d,c_3|a_2,b_3,a_1|c_5)
      W(d|a_1,a_3,a_2|c_4,c_5,c_6|b_4) .
\label{2.2}
\end{eqnarray}

The Boltzmann weight of the original Bazhanov-Baxter
form, which we denote by $W_0$ instead of $W$, is given by

\begin{eqnarray}
 \hskip -10mm W_0(a|e,f,g |b,c,d|h) =
   \sum^{N}_{\sigma=1}
    \frac{w(x_{3},x_{13},x_{1}|d,h + \sigma)
    w(x_{4},x_{24},x_{2}|a,g+ \sigma)}
   {w(x_{4},x_{14},x_{1}|e,c+ \sigma)
    w(x_{3}/\omega ,x_{23},x_{2}|f,b+ \sigma)},
\label{2.3}
\end{eqnarray}

\noindent
and $w(x,y,z|k,l)$ is given by

\begin{eqnarray}
  &&w(x,y,z|k,l)=\Phi(l)w(x,y,z|k-l)
   =\Phi(l) \prod_{s=1}^{k-l} {y \over {z-x \omega^{s} }}
  = \Phi(l) (y/z)^{k-l} w(x/z|k-l),
  \nonumber \\
  &&{\rm where} \quad \Phi(l)=\omega^{l(l+N)/2},
   \omega=e^{2 \pi i/N}.
   \nonumber
\end{eqnarray}

We impose the Fermat condition $x^N+y^N=z^N$ on $w(x,y,z|k,l)$ to
make this function $w(x,y,z|k,l)$ periodic under $k \rightarrow k+N$
and $l \rightarrow l+N$.

If the above Boltzmann weight satisfies the tetrahedron equation,
the simple Boltzmann weight

\begin{eqnarray}
  && \hskip -20mm
   W_1(a|e,f,g |b,c,d|h) \nonumber \\
  &&\hskip -16mm
   =\sum^{N}_{\sigma=1} \omega^{\sigma (-b-c+g+h)}
    \frac{w(x_{3},x_{13},x_{1}|d-h- \sigma)
    w(x_{4},x_{24},x_{2}|a-g- \sigma)}
   {w(x_{4},x_{14},x_{1}|e-c- \sigma)
    w(x_{3}/\omega ,x_{23},x_{2}|f-b- \sigma)}
\label{2.4}
\end{eqnarray}

\noindent
also satisfies the tetrahedron equation, as can be shown with a
simple calculation, and we use this form in this paper.

We also use the notation

\begin{eqnarray}
    &&w(v,l)=\Delta^{l}(v) \prod^{l}_{j=1}\frac{1}{1-v \omega^j},
 \nonumber\\
  && \hskip -48mm  {\rm where}\hskip 39mm  \Delta(v)=(1-v^N)^{1/N}.
   \label{2.5}
 \end{eqnarray}

\noindent
This function $w(v,l)$ automatically satisfies the periodicity
 condition $w(v,l+N)=w(v,l)$.
The connection between $w(x,y,z|l)$ and $w(v,l)$ is
given by

\begin{eqnarray}
 \hskip -10mm w(x,y,z|l)=(y/z)^{l} w(x/z|l)
   =\Delta^{l}(x/z) w(x/z|l)=w(x/z,l).  \nonumber
\end{eqnarray}

Using this function $w(v,l)$, we rewrite the Boltzmann weight of
Eq.~(\ref{2.4}) into the form

\begin{eqnarray}
  \hskip -10mm
  W_1(a|e,f,g |b,c,d|h) =
   \sum^{N}_{\sigma=1} \omega^{\sigma(-b-c+g+h)}
    \frac{w(v_3,d-h-\sigma)
    w(\tilde{v}_1,a-g-\sigma)}
   {w(v_4,e-c-\sigma)
    w(\tilde{v}_2,f-b-\sigma)},
\label{2.6}
\end{eqnarray}

\noindent
where $v_3=x_3/x_1$, $\tilde{v}_1=x_4/x_2$,
$v_4=x_4/x_1$, $\tilde{v}_2=x_3/\omega x_2$.

Later, we use the formula

\begin{eqnarray}
    w(v,-a)=\frac{1}{\Phi(a) w(\tilde{v},a)}, \quad
  {\rm where} \quad \tilde{v}=\frac{1}{\omega v}.
\label{2.7}
\end{eqnarray}

\subsection{The Bazhanov-Baxter solution of the tetrahedron equation}

By direct calculation, \cite{Kashaev2,ours1} it has been shown
that the conditions

\begin{eqnarray}
    \frac{x_2}{x_1}=\frac{x'_2}{x'_1} &,&
    \frac{x_{12}}{x_1}=\frac{x'_{12}}{x'_1},
\label{2.8}\\
    \frac{x_3}{\omega x_4}=\frac{x'''_2}{x'''_1} &,&
    \frac{x_{34}}{\omega^{1/2}x_4}=\frac{x'''_{12}}{x'''_1},
\label{2.9}\\
    \frac{x_{13}x_{24}}{x_{14}x_{23}}=\frac{x''_1}{x''_2} &,&
    \frac{x_{12}x_{34}}{x_{14}x_{23}}=\frac{x''_{12}}{x''_2},
\label{2.10}\\
    \frac{x'_{14}x'_{23}}{x'_{13}x'_{24}}
     =\frac{x''_{14}x''_{23}}{x''_{13}x''_{24}} &,&
    \frac{x'_{12}x'_{34}}{x'_{13}x'_{24}}
    =\frac{x''_{12}x''_{34}}{x''_{13}x''_{24}},
\label{2.11}\\
    \frac{x''_3}{x''_4}=\frac{x'''_3}{x'''_4} &,&
    \frac{x''_{34}}{x''_4}=\frac{x'''_{34}}{x'''_4},
\label{2.12}\\
    \frac{x'_4}{x'_3}=\frac{x'''_{13}x'''_{24}}
                     {\omega x'''_{14}x'''_{23}}&,&
    \frac{x'_{34}}{x'_3}
    =\frac{x'''_{12}x'''_{34}}{\omega^{1/2} x'''_{14}x'''_{23}},
\label{2.13}\\
%
%
\omega \frac{x_{23}}{x_3}\frac{x'_4}{x'_{24}}
       \frac{x''_{24}}{x''_2}\frac{x'''_2}{x'''_{24}}&=&1 ,
\label{2.14}\\
       \frac{x_{13}}{x_1}\frac{x'_1}{x'_{14}}
       \frac{x''_{14}}{x''_1}\frac{x'''_1}{x'''_{14}}&=&1 ,
\label{2.15}\\
       \frac{x_{14}}{x_4}\frac{x'_4}{x'_{14}}
       \frac{x''_{14}}{x''_4}\frac{x'''_4}{x'''_{24}}&=&1 ,
\label{2.16}\\
       \frac{x_{13}}{x_3}\frac{x'_3}{x'_{13}}
       \frac{x''_{13}}{x''_1}\frac{x'''_2}{x'''_{23}}&=&1
\label{2.17}
\end{eqnarray}

\noindent
are sufficient for the Boltzmann weights
$W_1$, $W'_1$, $W''_1$, $W'''_1$ to
satisfy the tetrahedron equation.

Using the angles $\theta_1$, $\theta_2$ and $\theta_3$ and the arcs
$a_1$, $a_2$ and $a_3$ opposite to the angles of the spherical triangles,
we parametrize as $W_1=W_1(\theta_2, \theta_1, \theta_3)$.
Similarly, we parametrize as
$W'_1=W_1(\pi-\theta_6, \theta_1, \pi-\theta_4)$,
$W''_1=W_1(\theta_5, \pi-\theta_3, \pi-\theta_4)$ and
$W'''_1=W_1(\theta_5, \theta_2, \theta_6)$.

The angles $\theta_1$ - $\theta_6$ are not independent. Rather,
they must satisfy one constraint condition, \cite{Zamolodchikov}

\begin{eqnarray}
 \left| \begin{array}{cccc}
        1 & -\cos{\theta_1} & -\cos{\theta_2} & -\cos{\theta_6} \\
        -\cos{\theta_1} & 1 & -\cos{\theta_3} & -\cos{\theta_4} \\
        -\cos{\theta_2} & -\cos{\theta_3} & 1 & -\cos{\theta_5} \\
        -\cos{\theta_6} & -\cos{\theta_4} & -\cos{\theta_5} & 1
   \end{array} \right|=0,
\label{2.18}
\end{eqnarray}

\noindent
which comes from the condition that all four unit vectors
$\vec{n}_i \ (i=1$ - $4)$, which are perpendicular to the plane
on which the large circles lie in 3-dim
space are not independent but must satisfy
${\rm det}_{ij}(\vec{n}_i \cdot \vec{n}_j)=0$.

The explicit spherical trigonometry parametrization of
$W_1(\theta_2, \theta_1, \theta_3)$ for the Bazhanov-Baxter model
is given by \cite{B-B,Kashaev1}

\begin{eqnarray}
  &&\hskip -10mm
  x_1=\frac{\cos^{1/N}(\theta_1/2)}{ \sin^{1/N}(\theta_1/2)}, \;\;\;
  x_2= \frac{\sin^{1/N}(\theta_1/2)}{\omega^{1/2} \cos^{1/N}(\theta_1/2)},
\label{2.19}\\
  &&\hskip -10mm
  x_3=\frac{\exp(-i a_3/N) \sin^{1/N}(\theta_2/2)}
           {\cos^{1/N}(\theta_2/2)}, \;\;\;
  x_4=\frac{\exp(-i a_3/N)
    \cos^{1/N}(\theta_2/2)}{\omega^{1/2} \sin^{1/N}(\theta_2/2)},
\label{2.20}\\
  &&\hskip -10mm
  x_{12}=\frac{1}{\cos^{1/N}(\theta_1/2) \sin^{1/N}(\theta_1/2)},\;\;\;
  x_{13}=\frac{\exp(i \beta_3/N) \sin^{1/N}(\theta_3/2)}
          { \sin^{1/N}(\theta_1/2) \cos^{1/N}(\theta_2/2)} ,
\label{2.21}\\
  &&\hskip -10mm
  x_{14}=\frac{\exp(-i \beta_2/N) \cos^{1/N}(\theta_3/2)}
         {\sin^{1/N}(\theta_1/2) \sin^{1/N}(\theta_2/2)},  \;\;\;
  x_{23}=\frac{\exp(-i \beta_1/N) \cos^{1/N}(\theta_3/2)}
      {\omega^{1/2} \cos^{1/N}(\theta_1/2) \cos^{1/N}(\theta_2/2)},
\label{2.22}\\
  &&\hskip -10mm
  x_{24}=\frac{\exp(i \beta_0/N) \sin^{1/N}(\theta_3/2)}
     {\omega^{1/2} \sin^{1/N}(\theta_2/2) \cos^{1/N}(\theta_1/2)}, \;\;\;
  x_{34}=\frac{\exp(-i a_3/N ) }{\cos^{1/N}(\theta_2/2)
         \sin^{1/N}(\theta_2/2)} ,
\label{2.23}
\end{eqnarray}

\noindent
where $\beta_1=(-a_1+a_2+a_3)/2$, $\beta_1=(a_1-a_2+a_3)/2$,
$\beta_3=(a_1+a_2-a_3)/2$ and $\beta_0=(2 \pi-a_1-a_2-a_3)/2$.

The parametrization of $W'_1$, $W''_1$ and $W'''_1$ is obtained from
the parametrization of $W_1$ by the replacements
$W'_1=W_1(\pi-\theta_6, \theta_1, \pi-\theta_4)$,
$W''_1=W_1(\theta_5, \pi-\theta_3, \pi-\theta_4)$ and
$W'''_1=W_1(\theta_5, \theta_2, \theta_6)$.

\subsection{The cubic symmetries of the Bazhanov-Baxter model}

Cubic symmetries are one of the essential points to
understand the structure of the Bazhanov-Baxter model.
We clarify the cubic symmetries in such a way as to give
a guiding principle to find a candidate as a solution
of the tetrahedron equation. Then, without restricting to the
Bazhanov-Baxter model, we investigate the kind of relations that
arise if we impose the cubic symmetries for the model.

For the cube, on which the Boltzmann weight is assigned, we attach
a spherical triangle with angles $\theta_1$, $\theta_2$ and
$\theta_3$.
We denote by $a_1$, $a_2$ and $a_3$  the arcs of the spherical
triangle opposite to the angles $\theta_1$, $\theta_2$ and $\theta_3$.
We denote the Boltzmann weight of this cube as
$W_1(a|e,f,g |b,c,d|h; \theta_2, \theta_1, \theta_3)$.


\vspace{6mm}

\unitlength 3pt
\begin{picture}
(40,40)
\put(51,11){$e$}           \put(71,11){$d$}
\put(39,20){$c$}           \put(55,15){$h$}
\put(51,27){$a$}           \put(72,30){$f$}
\put(39,41){$g$}           \put(59,40){$b$}

\bezier{100}(49,39)(52,37)(55,35)
\bezier{200}(55,35)(61,31)(62,20)
\bezier{100}(62,20)(62,15)(62,10)

\bezier{100}(38,28)(41,26)(44,24)
\bezier{200}(44,24)(48,22)(60,22)
\bezier{100}(60,22)(65,22)(70,22)

\bezier{100}(42,16)(42,20)(42,25)
\bezier{200}(42,25)(42,36)(52,35)
\bezier{100}(52,35)(57,35)(63,35)

\bezier{80}(42,29)(43,29)(44,24)
\bezier{80}(52,35)(54,33)(57,33)
\bezier{80}(59,22)(60,25)(61,25)

\put (44,26){$\theta_2$}
\put (52,31){$\theta_1$}
\put (55,24){$\theta_3$}
\put (50,10){\circle*{2}}
\put (70,10){\circle*{2}}
\put (38,19){\circle*{2}}
\put (58,19){\circle*{2}}
\put (50,30){\circle*{2}}
\put (70,30){\circle*{2}}
\put (38,39){\circle*{2}}
\put (58,39){\circle*{2}}

\thicklines
\put(50,10){\line(0,1){20}}
\put(70,10){\line(0,1){20}}
\put(38,19){\line(0,1){20}}
\put(50,10){\line(1,0){20}}
\put(50,30){\line(1,0){20}}
\put(38,39){\line(1,0){20}}
\put(50,10){\line(-4,3){12}}
\put(50,30){\line(-4,3){12}}
\put(70,30){\line(-4,3){12}}
\thinlines
\put(58,19){\line(0,1){20}}
\put(38,19){\line(1,0){20}}
\put(70,10){\line(-4,3){12}}
\put(30,2){{\bf Fig.~2}~ Spherical triangle for the cube}

\end{picture}

\vspace{4mm}

The cubic symmetries are composed of the following $\rho $-
and $\tau $-symmetries:

\begin{eqnarray}
  \rho-{\rm symmetry}:
    a,b,c,d,e,f,g,h
           &\rightarrow & g,f,h,e,c,a,b,d ,  \nonumber\\
    \theta_1, \theta_2, \theta_3
    &\rightarrow & \pi-\theta_1, \theta_3, \pi-\theta_2, \nonumber\\
    a_1, a_2, a_3  &\rightarrow & \pi-a_1, a_3, \pi-a_2,   
\label{2.24} \\
           && \nonumber\\
  \tau-{\rm symmetry}:
   a,b,c,d,e,f,g,h
         &\rightarrow & a,c,b,d,f,e,g,h ,   \nonumber\\
   \theta_1, \theta_2, \theta_3,
           &\rightarrow & \theta_2, \theta_1, \theta_3, \nonumber\\
   a_1, a_2, a_3,
           &\rightarrow &  a_2, a_1, a_3.  
\label{2.25}
\end{eqnarray}

\vspace{4mm}

\unitlength 3pt
\begin{picture}
(150,43)
(20,0)

\put(51,11){$e$}           \put(71,11){$d$}
\put(39,20){$c$}           \put(60,20){$h$}
\put(51,31){$a$}           \put(71,31){$f$}
\put(39,41){$g$}           \put(59,40){$b$}

\put (50,10){\circle*{2}}
\put (70,10){\circle*{2}}
\put (38,19){\circle*{2}}
\put (58,19){\circle*{2}}
\put (50,30){\circle*{2}}
\put (70,30){\circle*{2}}
\put (38,39){\circle*{2}}
\put (58,39){\circle*{2}}

\thicklines
\put(50,10){\line(0,1){20}}
\put(70,10){\line(0,1){20}}
\put(38,19){\line(0,1){20}}
\put(50,10){\line(1,0){20}}
\put(50,30){\line(1,0){20}}
\put(38,39){\line(1,0){20}}
\put(50,10){\line(-4,3){12}}
\put(50,30){\line(-4,3){12}}
\put(70,30){\line(-4,3){12}}
\thinlines
\put(58,19){\line(0,1){20}}
\put(38,19){\line(1,0){20}}
\put(70,10){\line(-4,3){12}}

\put(101,11){$c$}           \put(121,11){$e$}
\put(89,20){$h$}           \put(110,20){$d$}
\put(101,32){$g$}           \put(121,31){$a$}
\put(89,40){$b$}           \put(109,40){$f$}

\put (100,10){\circle*{2}}
\put (120,10){\circle*{2}}
\put (88,19){\circle*{2}}
\put (108,19){\circle*{2}}
\put (100,30){\circle*{2}}
\put (120,30){\circle*{2}}
\put (88,39){\circle*{2}}
\put (108,39){\circle*{2}}

\thicklines
\put(100,10){\line(0,1){20}}
\put(120,10){\line(0,1){20}}
\put(88,19){\line(0,1){20}}
\put(100,10){\line(1,0){20}}
\put(100,30){\line(1,0){20}}
\put(88,39){\line(1,0){20}}
\put(100,10){\line(-4,3){12}}
\put(100,30){\line(-4,3){12}}
\put(120,30){\line(-4,3){12}}
\thinlines
\put(108,19){\line(0,1){20}}
\put(88,19){\line(1,0){20}}
\put(120,10){\line(-4,3){12}}

\put(75,26){\vector(1,0){10}}
\put(77,28){ $\rho $}

\put(60,2){{\bf Fig.~3}~ $\rho $-symmetry for the cube}

\end{picture}

\vspace{4mm}

\unitlength 3pt
\begin{picture}
(150,43)
(20,0)
\put(51,11){$e$}           \put(71,11){$d$}
\put(39,20){$c$}           \put(60,20){$h$}
\put(51,31){$a$}           \put(71,31){$f$}
\put(39,41){$g$}           \put(59,40){$b$}

\put (50,10){\circle*{2}}
\put (70,10){\circle*{2}}
\put (38,19){\circle*{2}}
\put (58,19){\circle*{2}}
\put (50,30){\circle*{2}}
\put (70,30){\circle*{2}}
\put (38,39){\circle*{2}}
\put (58,39){\circle*{2}}

\thicklines
\put(50,10){\line(0,1){20}}
\put(70,10){\line(0,1){20}}
\put(38,19){\line(0,1){20}}
\put(50,10){\line(1,0){20}}
\put(50,30){\line(1,0){20}}
\put(38,39){\line(1,0){20}}
\put(50,10){\line(-4,3){12}}
\put(50,30){\line(-4,3){12}}
\put(70,30){\line(-4,3){12}}
\thinlines
\put(58,19){\line(0,1){20}}
\put(38,19){\line(1,0){20}}
\put(70,10){\line(-4,3){12}}

\put(101,11){$f$}           \put(121,11){$d$}
\put(89,20){$b$}           \put(110,20){$h$}
\put(101,31){$a$}           \put(121,31){$e$}
\put(89,41){$g$}           \put(109,40){$c$}

\put (100,10){\circle*{2}}
\put (120,10){\circle*{2}}
\put (88,19){\circle*{2}}
\put (108,19){\circle*{2}}
\put (100,30){\circle*{2}}
\put (120,30){\circle*{2}}
\put (88,39){\circle*{2}}
\put (108,39){\circle*{2}}

\thicklines
\put(100,10){\line(0,1){20}}
\put(120,10){\line(0,1){20}}
\put(88,19){\line(0,1){20}}
\put(100,10){\line(1,0){20}}
\put(100,30){\line(1,0){20}}
\put(88,39){\line(1,0){20}}
\put(100,10){\line(-4,3){12}}
\put(100,30){\line(-4,3){12}}
\put(120,30){\line(-4,3){12}}
\thinlines
\put(108,19){\line(0,1){20}}
\put(88,19){\line(1,0){20}}
\put(120,10){\line(-4,3){12}}

\put(75,26){\vector(1,0){10}}
\put(77,28){ $\tau $}

\put(60,2){{\bf Fig.~4}~ $\tau $-symmetry for the cube}

\end{picture}

\vspace{4mm}

Strictly speaking, in order that the $\rho$-symmetry exists, we must
multiply the original Boltzmann weight by the proper external spin dependent
prefactor. The tetrahedron equation is satisfied for the prefactor
and for the original Boltzmann weight separately.

In Eqs.~(\ref{2.24}) and (\ref{2.25}), we  have used the following
spherical trigonometry relations in order to know how
$a_1$, $a_2$ and $a_3$ change under the discrete
change of the $\theta_1$, $\theta_2$ and $\theta_3$:

\begin{eqnarray}
  &&\cos(a_i/2)
   =\sqrt{\frac{\cos \left\{(\theta_j-\theta_k+\theta_i)/2 \right\}
      \cos \left\{(-\theta_j+\theta_k+\theta_i)/2 \right\}}
   {\sin{\theta_j} \sin{\theta_k}}},
\label{2.26}\\
  &&\sin(a_i/2)
   =\sqrt{\frac{-\cos \left\{(\theta_j+\theta_k+\theta_i)/2 \right\}
      \cos \left\{(\theta_j+\theta_k-\theta_i)/2 \right\}}
   {\sin{\theta_j} \sin{\theta_k}}}.
\label{2.27}\\
     &&\qquad \qquad  \ (i \ne j \ne k \ne i =1, 2, 3)  \nonumber
\end{eqnarray}

\noindent
These come from the fundamental relations of
the spherical trigonometry,

\begin{eqnarray}
 &&\hskip -6mm
     \cos{\theta_i}=-\cos{\theta_j} \cos{\theta_k}+\cos{a_i}
     \sin{\theta_j} \sin{\theta_k},
     \ (i \ne j \ne k \ne i =1, 2, 3)
\label{2.28}\\
  &&\hskip -6mm
     \cos{a_i}=\cos{a_j} \cos{a_k}+\cos{\theta_i}
     \sin{a_j} \sin{a_k},
     \ (i \ne j \ne k \ne i =1, 2, 3)
\label{2.29}\\
 &&\hskip -6mm
    \frac{\sin{\theta_1}}{\sin{a_1}}
    =\frac{\sin{\theta_2}}{\sin{a_2}}
    =\frac{\sin{\theta_3}}{\sin{a_3}}.
\label{2.30}
\end{eqnarray}

If the Boltzmann weight has the above cubic symmetries,
we can rewrite the tetrahedron equation into the
form \cite{Baxter}

\begin{eqnarray}
   \sum^{N}_{d=1} &&
    W_1(a_4|c_2,c_3,c_1|b_1,b_2,b_3|d)
    W'_1(a_3|c_1,c_6,c_4|b_4,b_1,b_2|d)
                   \nonumber \\
   &&\times W''_1(a_2|c_4,c_3,c_5|b_3,b_4,b_1|d)
   W'''_1(a_1|c_5,c_6,c_2|b_2,b_3,b_4|d)
                            \nonumber \\
 =\sum^{N}_{d=1}&&
   W'''_1(b_1|c_1,c_3,c_4|a_2,a_3,a_4|d)
   W''_1(b_2|c_2,c_6,c_1|a_3,a_4,a_1|d)             \nonumber \\
   &&  \times  W'_1(b_3|c_5,c_3,c_2|a_4,a_1,a_2|d)
      W_1(b_4|c_4,c_6,c_5|a_1,a_2,a_3|d).
\label{2.31}
\end{eqnarray}

\noindent
Then if we take $a_1=b_1$, $a_2=b_2$, $a_3=b_3$, $a_4=b_4$,
$c_1=c_5$, $c_2=c_4$ and $c_3=c_6$, the tetrahedron equation
is automatically satisfied; that is, if the cubic
symmetries exist, $N^7$ of the $N^{14}$ relations in the
tetrahedron equation are automatically satisfied, so that if
the cubic symmetries exist, it is quite likely that the
tetrahedron equation is satisfied.
Of course, there may exist a solution of the
tetrahedron equation without the cubic symmetries.

The full cubic symmetries, especially $\rho $-symmetry,
are too restrictive, and it seems difficult to find
a full cubic symmetric solution other than the Bazhanov-Baxter
model. Thus we use a part of the cubic symmetries to give a
guiding principle to find a candidate as a new solution.
These symmetries have the properties $\rho^4=1$, $ \tau^2=1$,
$(\rho \tau)^6=1$, and we use the special $\rho^2$, $ \tau $ and
$(\rho \tau)^3$ cubic symmetries here.

\noindent
i) {\it  $\rho^2$-symmetry} \hfill

Under $\rho^2$-symmetry, the spin variables, the angles
and the arcs change as

\begin{eqnarray}
  a,b,c,d,e,f,g,h &\rightarrow & b,a,d,c,h,g,f,e, \nonumber \\
    \theta_1, \theta_2, \theta_3
  &\rightarrow & \theta_1, \pi-\theta_2, \pi-\theta_3
\nonumber \\
    a_1, a_2, a_3 &\rightarrow & a_1, \pi-a_2, \pi-a_3,
\label{2.32}
\end{eqnarray}

\noindent
and the Boltzmann weight changes as

\begin{eqnarray}
  &&W_1(a|e,f,g |b,c,d|h;\theta_2,\theta_1, \theta_3) \rightarrow
   \sum^{N}_{\sigma=1} \omega^{\sigma(-a-d+f+e)}
    \frac{w(v'_3,c-e-\sigma)
    w(\tilde{v}'_1,b-f-\sigma)}
   {w(v'_4,h-d-\sigma)
    w(\tilde{v}'_2,g-a-\sigma)} \nonumber \\
   &&={\rm (const)}
   \sum^{N}_{\sigma=1} \omega^{\sigma(-b-c+g+h)}
    \frac{w(\tilde{v}'_4,d-h-\sigma)
    w(v'_2,a-g-\sigma)}
   {w(\tilde{v}'_3,e-c-\sigma)
    w(v'_1,f-b-\sigma)} ,
\label{2.33}
\end{eqnarray}

\noindent
where we have used Eq.~(\ref{2.7}). If we impose the cubic symmetries
for the model, this transformed Boltzmann weight must be
proportional to the original Boltzmann weight, that is,

\begin{eqnarray}
  &&\hskip -6mm
   v'_2=v_2(\pi-\theta_2, \theta_1, \pi-\theta_3)=
    \frac{1}{\omega \tilde{v}_2(\pi-\theta_2, \theta_1, \pi-\theta_3)}
    =\tilde{v}_1(\theta_2, \theta_1, \theta_3) ,
\label{2.34}\\
  &&\hskip -6mm
    v'_1=v_1(\pi-\theta_2, \theta_1, \pi-\theta_3)=
    \frac{1}{\omega \tilde{v}_1(\pi-\theta_2, \theta_1, \pi-\theta_3)}
    =\tilde{v}_2(\theta_2, \theta_1, \theta_3)  ,
\label{2.35}\\
  &&\hskip -6mm
    \tilde{v}'_4=\tilde{v}_4(\pi-\theta_2, \theta_1, \pi-\theta_3)=
    \frac{1}{\omega v_4(\pi-\theta_2, \theta_1, \pi-\theta_3)}
    =v_3(\theta_2, \theta_1, \theta_3) ,
\label{2.36}\\
  &&\hskip -6mm
   \tilde{v}'_3=\tilde{v}_3(\pi-\theta_2, \theta_1, \pi-\theta_3)=
    \frac{1}{\omega v_3(\pi-\theta_2, \theta_1, \pi-\theta_3)}
    =v_4(\theta_2, \theta_1, \theta_3) .
\label{2.37}
\end{eqnarray}

\noindent
These give

\begin{eqnarray}
   && \omega \tilde{v}_1(\theta_2, \theta_1, \theta_3)
    \tilde{v}_2(\pi-\theta_2, \theta_1, \pi-\theta_3)=1,
\label{2.38}\\
      && \omega v_3(\theta_2, \theta_1, \theta_3)
    v_4(\pi-\theta_2, \theta_1, \pi-\theta_3)=1.
\label{2.39}
\end{eqnarray}

\noindent
ii) {\it $\tau$-symmentry} \hfill

Under $\tau $-symmetry, the spin variables, the angles and
the arcs change as

\begin{eqnarray}
   a,b,c,d,e,f,g,h &\rightarrow & a,c,b,d,f,e,g,h , \nonumber \\
   \theta_1, \theta_2, \theta_3   &\rightarrow &
   \theta_2,  \theta_1, \theta_3,
\nonumber \\
   a_1, a_2, a_3   &\rightarrow & a_2, a_1, a_3,
   \label{2.40}
\end{eqnarray}

\noindent
and the Boltzmann weight changes according to

\begin{eqnarray}
  &&\hskip -45 mm W_1(a|e,f,g |b,c,d|h;\theta_2,\theta_1, \theta_3) 
     \nonumber \\
  && \hskip -40 mm \rightarrow
   \sum^{N}_{\sigma=1} \omega^{\sigma(-b-c+g+h)}
    \frac{w(v''_3,d-h-\sigma)
    w(\tilde{v}''_1,a-g-\sigma)}
   {w(v''_4,f-b-\sigma)
    w(\tilde{v}''_2,e-c-\sigma)}.
\label{2.41}
\end{eqnarray}

If we impose cubic symmetries for the model, this
transformed Boltzmann weight
must be proportional to the original Boltzmann weight; that is,

\begin{eqnarray}
  &&v''_3=v_3(\theta_1, \theta_2, \theta_3)=
    v_3(\theta_2, \theta_1, \theta_3) ,
\label{2.41i}\\
 &&\tilde{v}''_1=\tilde{v}_1(\theta_1, \theta_2, \theta_3)=
    \tilde{v}_1(\theta_2, \theta_1, \theta_3) ,
\label{2.41ii}\\
  &&v''_4=v_4(\theta_1, \theta_2, \theta_3)=
    \tilde{v}_2(\theta_2, \theta_1, \theta_3) ,
\label{2.42}\\
  &&\tilde{v}''_2=\tilde{v}_2(\theta_1, \theta_2, \theta_3)=
    v_4(\theta_2, \theta_1, \theta_3) ,
\label{2.43}
\end{eqnarray}

\noindent
which give

\begin{eqnarray}
   &&v_4(\theta_1, \theta_2, \theta_3)=
    \tilde{v}_2(\theta_2, \theta_1, \theta_3),
\label{2.44}\\
   && v_3(\theta_1, \theta_2, \theta_3)=
    v_3(\theta_2, \theta_1, \theta_3),
\label{2.44i}\\
   && \tilde{v}_1(\theta_1, \theta_2, \theta_3)=
    \tilde{v}_1(\theta_2, \theta_1, \theta_3).
\label{2.44ii}
\end{eqnarray}

Using Eqs.~(\ref{2.38}), (\ref{2.39}), (\ref{2.44}),
(\ref{2.44i}) and (\ref{2.44ii}), we can write
$\tilde{v}_1$, $\tilde{v}_2$, $v_4$ with $v_3$ in the
form

\begin{eqnarray}
   &&\tilde{v}_1(\theta_2,\theta_1, \theta_3)
   =v_3(\pi-\theta_1,\pi-\theta_2, \theta_3) ,
\label{2.45}\\
   &&\tilde{v}_2(\theta_2,\theta_1, \theta_3)
   =\frac{1}{\omega v_3(\pi-\theta_1, \theta_2, \pi-\theta_3)} ,
\label{2.46}\\
   &&v_4(\theta_2,\theta_1, \theta_3)
   =\frac{1}{\omega v_3(\pi-\theta_2, \theta_1, \pi-\theta_3)},
\label{2.47}
\end{eqnarray}

\noindent
where $v_3(\theta_1, \theta_2, \theta_3)$$=$
$v_3(\theta_2, \theta_1, \theta_3)$.


\noindent
iii) {\it $(\rho\tau)^3$-symmentry} \hfill

Under $(\rho \tau)^3$-symmetry, the spin variables, the angles
and the arcs change as

\begin{eqnarray}
   a,b,c,d,e,f,g,h &\rightarrow & h,e,f,g,b,c,d,a , \nonumber \\
    \theta_1, \theta_2, \theta_3   &\rightarrow &
   \theta_1, \theta_2, \theta_3,
\nonumber \\
    a_1, a_2, a_3   &\rightarrow & a_1, a_2, a_3,   \label{2.48}
\end{eqnarray}
\noindent
and the Boltzmann weight changes according to

\begin{eqnarray}
  &&W_1(a|e,f,g |b,c,d|h; \theta_2, \theta_1, \theta_3)
    \rightarrow
   \sum^{N}_{\sigma=1} \omega^{\sigma(-e-f+a+d)}
    \frac{w(v_3,g-a-\sigma)
    w(\tilde{v}_1,h-d-\sigma)}
   {w(v_4,b-f-\sigma)
    w(\tilde{v}_2,c-e-\sigma)}  \nonumber \\
  &&\;\;\;\;\;={\rm (const)}
   \sum_{\sigma \in Z_{N}} \omega^{\sigma(-b-c+g+h)}
    \frac{w(\bar{v}_3,d-h-\sigma)
    w(\tilde{\bar{v}}_1,a-g-\sigma)}
   {w(\bar{v}_4,e-c-\sigma)
    w(\tilde{\bar{v}}_2,f-b-\sigma)} .
\label{2.49}
\end{eqnarray}

\noindent

In the above, we have used the Star-Star relation of Eqs.~(A.17) and
 (A.18) in the paper of Sergeev et al.,\cite{Mangazeev2}
and we have identified the spin variables $m_1=h-d$, $m_2=c-e$, $m_3=g-a$,
$m_4=b-f$, $n=-(-e-f+a+d)$, $\bar{n}=n-m_1-m_3+m_2+m_4=-(-b-c+g+h)$
in their formula. In the above, we can represent $\bar{v}_i$ as the function
of $v_i$ but we do not give the explicit form, as it is quite
complicated. If we impose the cubic symmetries for the model,
the last expression of Eq.~(\ref{2.49}) must be proportional to
the original Boltzmann weight; that is,
$\bar{v}_i=v_i$, $(i=1 \sim 4)$, which becomes
equivalent to the condition \cite{Mangazeev2}

\begin{eqnarray}
   \omega \tilde{v}_2(\theta_2, \theta_1, \theta_3)
     v_4(\theta_2, \theta_1, \theta_3)
   =\tilde{v}_1(\theta_2, \theta_1, \theta_3)
     v_3(\theta_2, \theta_1, \theta_3).
   \nonumber
\end{eqnarray}

Substituting Eqs.~(\ref{2.45}) - (\ref{2.47}) into
the above, we have

\begin{eqnarray}
  && \omega v_3(\theta_1, \theta_2, \theta_3)
    v_3(\pi-\theta_1, \pi-\theta_2, \theta_3)
     \nonumber \\
  && \times
    v_3(\theta_1, \pi-\theta_2, \pi-\theta_3)
      v_3(\pi-\theta_1, \theta_2, \pi-\theta_3)=1,
\label{2.50}
\end{eqnarray}

\noindent
where $v_3(\theta_1, \theta_2, \theta_3)$$=$
$v_3(\theta_2, \theta_1, \theta_3)$.

In this way, even a part of the cubic symmetries gives a
strong constraint on the functional forms of
$\tilde{v}_1$, $\tilde{v}_2$, $v_3$ and $v_4$.

\subsection{Spherical trigonometry parametrization (I)
-approach using cubic symmetries  -}

In addition to the cubic symmetries, the spherical trigonometry
parametrization is also an essential point to understand the
structure of the Bazhanov-Baxter model.

Here, we give the approach using cubic symmetries
for the spherical trigonometry parametrization, which will give the
principle to find the candidate of the solution of the
tetrahedron equation.

We can prove the following spherical trigonometry relation
from Eqs.~(\ref{2.28}) - (\ref{2.30}):

\begin{eqnarray}
  && \cos(\theta_1/2) \cos(\theta_2/2)-\exp(-i a_3) 
  \sin(\theta_1/2) \sin(\theta_2/2) \nonumber \\
   && =\exp[i(a_1+a_2-a_3)/2] \sin(\theta_3/2) .
\label{2.51}
\end{eqnarray}

Rewriting this relation into the form

\begin{eqnarray}
 \hskip -10mm
  \frac{\exp(-i a_3) \sin(\theta_1/2)\sin(\theta_2/2)}
       {\cos(\theta_1/2) \cos(\theta_2/2)}
  +\frac{\exp[i(a_1+a_2-a_3)/2] \sin(\theta_3/2)}
        {\cos(\theta_1/2) \cos(\theta_2/2)}=1,
\label{2.52}
\end{eqnarray}

\noindent
and noting the relation $v^N + \Delta^{N}(v)=1$,
we have the parametrization

\begin{eqnarray}
  &&v_3(\theta_2, \theta_1, \theta_3)=\frac{x_3}{x_1}
   =\frac{\exp(-i a_3/N)
    \sin^{1/N}(\theta_1/2) \sin^{1/N}(\theta_2/2)}
       {\cos^{1/N}(\theta_1/2) \cos^{1/N}(\theta_2/2)} ,
\label{2.53}\\
  &&\Delta(v_3(\theta_2, \theta_1, \theta_3))
   =\frac{x_{13}}{x_1}
   = \frac{ \exp[i(a_1+a_2-a_3)/2N]
            \sin^{1/N}(\theta_3/2)}
       {\cos^{1/N}(\theta_1/2) \cos^{1/N}(\theta_2/2)},
\label{2.54}
\end{eqnarray}

\noindent
where we have taken the branch of the $N$-th root properly. This
functional form of $v_3(\theta_2,\theta_1, \theta_3)$ satisfies
the condition Eq.~(\ref{2.50}), which is the necessary condition
for the cubic symmetries to exist for the model.

Next, using the relation

\begin{eqnarray}
  &&\tilde{v}_1(\theta_2,\theta_1, \theta_3)
   =v_3(\pi-\theta_1,\pi-\theta_2, \theta_3), \nonumber \\
  &&\Delta(\tilde{v}_1(\theta_2,\theta_1, \theta_3))
   =\Delta(v_3(\pi-\theta_1,\pi-\theta_2, \theta_3)), \nonumber
\end{eqnarray}

\noindent
we have

\begin{eqnarray}
  &&\hskip -6mm
   \tilde{v}_1(\theta_2, \theta_1, \theta_3)
    =\frac{x_4}{x_2}
    =\frac{ \exp(-i a_3/N)
     \cos^{1/N}(\theta_1/2) \cos^{1/N}(\theta_2/2)}
       {\sin^{1/N}(\theta_1/2) \sin^{1/N}(\theta_2/2)}  ,
\label{2.55}\\
  &&\hskip -6mm
  \Delta(\tilde{v}_1(\theta_2, \theta_1, \theta_3))
  =\frac{x_{24}}{x_2}
  =\frac{ \exp[i(2 \pi-a_1-a_2-a_3)/2N]
        \sin^{1/N}(\theta_3/2)}
       {\sin^{1/N}(\theta_1/2) \sin^{1/N}(\theta_2/2)} ,
\label{2.56}
\end{eqnarray}

\noindent
where we have use the fact that $a_2 \rightarrow \pi-a_1$,
$a_1 \rightarrow \pi-a_2$, $a_3 \rightarrow a_3$
under
$\theta_2 \rightarrow \pi-\theta_1$,
$\theta_1 \rightarrow \pi-\theta_2$,
$\theta_3 \rightarrow \theta_3$.

Next, using the relation

\begin{eqnarray}
  &&\tilde{v}_2(\theta_2,\theta_1, \theta_3)
  =\frac{1}{\omega v_3(\pi-\theta_1, \theta_2, \pi-\theta_3)},
  \nonumber \\
  &&\Delta(\tilde{v}_2(\theta_2,\theta_1, \theta_3))
  = \frac{\Delta(v_3(\pi-\theta_1, \theta_2, \pi-\theta_3))}
  { \omega^{1/2} v_3(\pi-\theta_1, \theta_2, \pi-\theta_3)}
  \nonumber \\
  &&=\omega^{1/2} \tilde{v}_2(\theta_2,\theta_1, \theta_3)
   \Delta(v_3(\pi-\theta_1, \theta_2, \pi-\theta_3),  \nonumber
\end{eqnarray}

\noindent
we have

\begin{eqnarray}
  &&\hskip -6mm
   \tilde{v}_2(\theta_2, \theta_1, \theta_3)
   =\frac{x_3}{\omega x_2}
   =\frac{\exp(-i a_3/N)
            \cos^{1/N}(\theta_1/2) \sin^{1/N}(\theta_2/2)}
       {\omega^{1/2} \sin^{1/N}(\theta_1/2) \cos^{1/N}(\theta_2/2)} ,
\label{2.57}\\
  &&\hskip -6mm
   \Delta(\tilde{v}_2(\theta_2, \theta_1, \theta_3))
   =\frac{x_{23}}{x_2}
   =\frac{\exp[-i(-a_1+a_2+a_3)/2N]
      \cos^{1/N}(\theta_3/2)}
       {\sin^{1/N}(\theta_1/2) \cos^{1/N}(\theta_2/2)} ,
\label{2.58}
\end{eqnarray}

\noindent
where we have used the fact that $a_2 \rightarrow \pi-a_1$,
$a_1 \rightarrow a_2$, $a_3 \rightarrow \pi-a_3$
under $\theta_2 \rightarrow \pi-\theta_1$,
$\theta_1 \rightarrow \theta_2$,
$\theta_3 \rightarrow \pi-\theta_3$.

Next, using Eq.~(\ref{2.44}), we have
$v_4(\theta_2,\theta_1, \theta_3)
 =\tilde{v}_2(\theta_1,\theta_2, \theta_3)$, which gives

\begin{eqnarray}
  &&\hskip -6mm
   v_4(\theta_2, \theta_1, \theta_3)=\frac{x_4}{x_1}
   =  \frac{\exp(-i a_3/N)
    \sin^{1/N}(\theta_1/2) \cos^{1/N}(\theta_2/2)}
   {\omega^{1/2} \cos^{1/N}(\theta_1/2) \sin^{1/N}(\theta_2/2)} ,
\label{2.59}\\
  &&\hskip -6mm
    \Delta(v_4(\theta_2, \theta_1, \theta_3))=\frac{x_{14}}{x_1}
   =\frac{ \exp[-i(a_1-a_2+a_3)/2N]
    \cos^{1/N}(\theta_3/2)}
       {\cos^{1/N}(\theta_1/2) \sin^{1/N}(\theta_2/2)} .
\label{2.60}
\end{eqnarray}

In Eq.~(\ref{2.19}), we have chosen the normalization factor of $x_1$
in such a way that $x_1$ becomes
$x_1=\cos^{1/N}(\theta_1/2)/\sin^{1/N}(\theta_1/2)$.
Then
Eqs.~(\ref{2.19}) - (\ref{2.23}) give
the parametrization of the Bazhanov-Baxter model.

\subsection{Spherical trigonometry parametrization (II)
-intuitive approach-}

Here, we give the second approach, the intuitive approach,
for the spherical trigonometry parametrization.
In this approach, we do not assume any cubic symmetries for the model,
and we give a guiding principle to find a candidate as a solution
of the tetrahedron equation. In the next section, we give a new
solution of the tetrahedron equation, where we use this intuitive
approach.

We start from the same spherical trigonometry relation,
Eq.~(\ref{2.51}),

\begin{eqnarray}
  && \cos(\theta_1/2) \cos(\theta_2/2)- 
   \exp(-i a_3) \sin(\theta_1/2)
   \sin(\theta_2/2) \nonumber \\
   && =\exp[i(a_1+a_2-a_3)/2] \sin(\theta_3/2),
\label{2.61}
\end{eqnarray}

\noindent
and we parametrize as

\begin{eqnarray}
  &&x_1^N=A_1 \cos(\theta_1/2) \cos(\theta_2/2),
\label{2.62}\\
  &&x_3^N=A_1 \exp(-i a_3) \sin(\theta_1/2) \sin(\theta_2/2),
\label{2.63}\\
  &&x_{13}^N=A_1 \exp[i(a_1+a_2-a_3)/2] \sin(\theta_3/2) .
\label{2.64}
\end{eqnarray}

If there exists a spherical triangle with angles
$\{\theta_1$, $\theta_2$, $\theta_3\}$, there also exists
a spherical triangle with the angles $\{\pi-\theta_1$, $\pi-\theta_2$,
$\theta_3\}$.
Then, by replacing  $\theta_1 \rightarrow \pi-\theta_1$,
$\theta_2 \rightarrow \pi-\theta_2$,
$\theta_3 \rightarrow \theta_3$ in Eq.~(\ref{2.61}),
we have another form of the spherical trigonometry relation,

\begin{eqnarray}
  && \sin(\theta_1/2) \sin(\theta_2/2)
    -\exp(-i a_3) \cos(\theta_1/2)
   \cos(\theta_2/2) \nonumber \\
   && =\exp[i(2 \pi-a_1-a_2-a_3)/2] \sin(\theta_3/2).
\label{2.65}
\end{eqnarray}

Corresponding to the above relation, we parametrize as

\begin{eqnarray}
  &&x_2^N=A_2 \sin(\theta_1/2) \sin(\theta_2/2),
\label{2.66}\\
  &&x_4^N=A_2 \exp(-i a_3) \cos(\theta_1/2) \cos(\theta_2/2),
\label{2.67}\\
  &&x_{24}^N=A_2 \exp[i(2 \pi-a_1-a_2-a_3)/2] \sin(\theta_3/2) .
\label{2.68}
\end{eqnarray}

By replacing  $\theta_1 \rightarrow \theta_1$,
$\theta_2 \rightarrow \pi-\theta_2$,
$\theta_3 \rightarrow \pi-\theta_3$, and further, taking the complex
conjugate in Eq.~(\ref{2.61}),
we have another form of the spherical trigonometry relation

\begin{eqnarray}
  &&  \cos(\theta_1/2)\sin(\theta_2/2)+\exp(-i a_3)
   \sin(\theta_1/2) \cos(\theta_2/2) \nonumber \\
   && =\exp[-i(a_1-a_2+a_3)/2] \cos(\theta_3/2).
\label{2.69}
\end{eqnarray}

Using this relation, we parametrize as

\begin{eqnarray}
  &&x_1^N=A_3 \cos(\theta_1/2) \sin(\theta_2/2),
\label{2.70}\\
  &&x_4^N=-A_3 \exp(-i a_3) \sin(\theta_1/2) \cos(\theta_2/2),
\label{2.71}\\
  &&x_{14}^N=A_3 \exp[-i(a_1-a_2+a_3)/2] \sin(\theta_3/2) .
\label{2.72}
\end{eqnarray}

By replacing  $\theta_1 \rightarrow \pi-\theta_1$,
$\theta_2 \rightarrow \theta_2$,
$\theta_3 \rightarrow \pi-\theta_3$, and further, taking the complex
conjugate in Eq.~(\ref{2.61}),
we have another form of the spherical trigonometry relation,

\begin{eqnarray}
  &&  \sin(\theta_1/2)\cos(\theta_2/2)+\exp(-i a_3)
   \cos(\theta_1/2) \sin(\theta_2/2) \nonumber \\
   && =\exp[-i(-a_1+a_2+a_3)/2] \cos(\theta_3/2).
\label{2.73}
\end{eqnarray}

Using this relation, we parametrize as

\begin{eqnarray}
  &&x_2^N=A_4 \sin(\theta_1/2) \cos(\theta_2/2),
\label{2.74}\\
  &&x_3^N=-A_4 \exp(-i a_3) \cos(\theta_1/2) \sin(\theta_2/2),
\label{2.75}\\
  &&x_{23}^N=A_4 \exp[-i(-a_1+a_2+a_3)/2] \cos(\theta_3/2) .
\label{2.76}
\end{eqnarray}

If we take the overall factors to be
$A_1$$=$$1/\sin(\theta_1/2) \cos(\theta_2/2)$,
$A_2$$=$$-1/\cos(\theta_1/2) \sin(\theta_2/2)$,
$A_3$$=$$1/\sin(\theta_1/2) \sin(\theta_2/2)$ and
$A_4$$=$$-1/\cos(\theta_1/2) \cos(\theta_2/2)$,
the expressions of
$x^{N}_1$, $x^{N}_2$, $x^{N}_3$ and $x^{N}_4$ become consistent.
Taking the branch of the $N$-th root properly, we have
the parametrization of the Bazhanov-Baxter model.

\section{A new solution of the tetrahedron equation}

The spherical trigonometry relation Eq.~(\ref{2.51}) is the
spin $1/2$ representation in the $SU(2)$ language. We use one of the
fundamental relations of the spherical trigonometry relation
Eq.~(\ref{2.28}),

\begin{eqnarray}
  \cos{\theta_1} \cos{\theta_2}- \sin{\theta_1}
   \sin{\theta_2} \cos{a_3}=-\cos{\theta_3},
\label{3.1}
\end{eqnarray}

\noindent
which is the spin $1$ representation in the $SU(2)$ language.
Using this relation, we parametrize as

\begin{eqnarray}
  &&x_1^N=B_1 \cos{\theta_1} \cos{\theta_2},
\label{3.2}\\
  &&x_3^N=B_2  \sin{\theta_1} \sin{\theta_2} \cos{a_3},
\label{3.3}\\
  &&x_{13}^N=-B_1 \cos{\theta_3},
\label{3.4}
\end{eqnarray}

\noindent
and we have

\begin{eqnarray}
   v_3(\theta_2,\theta_1, \theta_3)=\frac{x_3}{x_1}
   =\frac{ \sin^{1/N}{\theta_1} \sin^{1/N}{\theta_2} \cos^{1/N}{a_3}}
    { \cos^{1/N}{\theta_1} \cos^{1/N}{\theta_2} }
\label{3.5}
\end{eqnarray}

If the cubic symmetries exist, $v_3$  must satisfies
the relation Eq.~(\ref{2.50}),
but the above $v_3$ does not satisfy this condition.
Then we must take a second intuitive approach to the spherical
trigonometry parametrization and attempt to find a candidate
as a solution of the tetrahedron equation. After we find the
candidate as a solution, we must check whether
it satisfies the condition of Kashaev et al.
Eqs.~(\ref{2.8}) - (\ref{2.17}).

If we change the angles as
$\{\theta_1 \rightarrow \pi-\theta_1$,
$\theta_2 \rightarrow \pi-\theta_2$,
$\theta_3 \rightarrow \theta_3\}$,
or
$\{\theta_1 \rightarrow \theta_1$,
$\theta_2 \rightarrow \pi-\theta_2$,
$\theta_3 \rightarrow \pi-\theta_3\}$,
or
$\{\theta_1 \rightarrow \pi-\theta_1$,
$\theta_2 \rightarrow \theta_2$,
$\theta_3 \rightarrow \pi-\theta_3\}$,
they give the same spherical trigonometry relation Eq.~(\ref{3.1}).
Then we parametrize as follows:

\begin{eqnarray}
  &&x_2^N=B_2 \cos{\theta_1} \cos{\theta_2},
\label{3.6}\\
  &&x_4^N=B_2  \sin{\theta_1} \sin{\theta_2} \cos{a_3},
\label{3.7}\\
  &&x_{24}^N=-B_2 \cos{\theta_3},
\label{3.8}\\
  &&x_1^N=B_3 \cos{\theta_1} \cos{\theta_2},
\label{3.9}\\
  &&x_4^N=B_3 \sin{\theta_1} \sin{\theta_2} \cos{a_3},
\label{3.10}\\
  &&x_{14}^N=-B_3 \cos{\theta_3} ,
\label{3.11}\\
  &&x_2^N=B_4 \cos{\theta_1} \cos{\theta_2},
\label{3.12}\\
  &&x_3^N=B_4 \sin{\theta_1} \sin{\theta_2} \cos{a_3},
\label{3.13}\\
  &&x_{23}^N=-B_4 \cos{\theta_3}.
\label{3.14}
\end{eqnarray}

We take the normalization factor to be $B_1=B_2=B_3=B_4=1$,
and we take the branch of the $N$-th root in such a way as the
$\{ x_i, x_{ij} \}$ satisfy the condition of Kashaev et al..

Then the  parametrization of $W_1=W_1(\theta_2, \theta_1, \theta_3)$
is given by

\begin{eqnarray}
 && x_1=x_2=\cos^{1/N}{\theta_1} \cos^{1/N}{\theta_2},
\label{3.15}\\
 && x_3=\omega x_4=\sin^{1/N}{\theta_1} \sin^{1/N}{\theta_2}
                   \cos^{1/N}{a_3},
\label{3.16}\\
 && x_{12}=x_{34}=0,
\label{3.17}\\
 && x_{13}=x_{14}=x_{23}=x_{24}=\omega^{1/2} \cos^{1/N}{\theta_3}.
\label{3.18}
\end{eqnarray}

The Boltzmann weights are parametrized by the angles of the
spherical triangle in the following way:
$W_1=W_1(\theta_2, \theta_1, \theta_3)$,
$W'_1=W_1(\pi-\theta_6, \theta_1, \pi-\theta_4)$,
$W''_1=W_1(\theta_5, \pi-\theta_3, \pi-\theta_4)$ and
$W'''_1=W_1(\theta_5, \theta_2, \theta_6)$, so that we can obtain
$\{x'_i, x'_{ij}\}$, $\{x''_i, x''_{ij}\}$ and $\{x'''_i, x'''_{ij}\}$ 
from $\{x_i, x_{ij}\}$ by the above replacement of angles.
The explicit form is given in the following form:

\begin{eqnarray}
 && x'_1=x'_2=\cos^{1/N}{\theta_1} \cos^{1/N}{(\pi-\theta_6)},
\label{3.19}\\
 && x'_3=\omega x'_4=\sin^{1/N}{\theta_1} \sin^{1/N}{(\pi-\theta_6)}
                   \cos^{1/N}{a'_3},
\label{3.20}\\
 && x'_{12}=x'_{34}=0,
\label{3.21}\\
 && x'_{13}=x'_{14}=x'_{23}=x'_{24}=\omega^{1/2}
  \cos^{1/N}{(\pi-\theta_4)},
\label{3.22}\\
 && x''_1=x''_2= \cos^{1/N}{(\pi-\theta_3)} \cos^{1/N}{\theta_5},
\label{3.23}\\
 && x''_3=\omega x''_4= \sin^{1/N}{(\pi-\theta_3)} \sin^{1/N}{\theta_5}
                   \cos^{1/N}{a''_3},
\label{3.24}\\
 && x''_{12}=x''_{34}=0,
\label{3.25}\\
 && x''_{13}=x''_{14}=x''_{23}=x''_{24}
     =\omega^{1/2} \cos^{1/N}{(\pi-\theta_4)},
\label{3.26}\\
 && x'''_1=x'''_2=\cos^{1/N}{\theta_2} \cos^{1/N}{\theta_5},
\label{3.27}\\
 && x'''_3=\omega x'''_4=\sin^{1/N}{\theta_2} \sin^{1/N}{\theta_5}
                   \cos^{1/N}{a'''_3},
\label{3.28}\\
 && x'''_{12}=x'''_{34}=0,
\label{3.29}\\
 && x'''_{13}=x'''_{14}=x'''_{23}=x'''_{24}
  =\omega^{1/2} \cos^{1/N}{\theta_6}.
\label{3.30}
\end{eqnarray}

Substituting these relations, we can show that
Eqs.~(\ref{2.8}) - (\ref{2.13}) are satisfied in the following way:

\begin{eqnarray}
    \frac{x_2}{x_1}=1=\frac{x'_2}{x'_1} &,&\;\;\;\;\;
    \frac{x_{12}}{x_1}=0=\frac{x'_{12}}{x'_1},
\label{3.31}\\
    \frac{x_3}{\omega x_4}=1=\frac{x'''_2}{x'''_1} &,&\;\;\;\;\;
    \frac{x_{34}}{\omega^{1/2}x_4}=0=\frac{x'''_{12}}{x'''_1},
\label{3.32}\\
    \frac{x_{13}x_{24}}{x_{14}x_{23}}=1=\frac{x''_1}{x''_2} &,&\;\;\;\;\;
    \frac{x_{12}x_{34}}{x_{14}x_{23}}=0=\frac{x''_{12}}{x''_2},
\label{3.33}\\
    \frac{x'_{14}x'_{23}}{x'_{13}x'_{24}}=1
     =\frac{x''_{14}x''_{23}}{x''_{13}x''_{24}} &,&\;\;\;\;\;
    \frac{x'_{12}x'_{34}}{x'_{13}x'_{24}}=0
    =\frac{x''_{12}x''_{34}}{x''_{13}x''_{24}},
\label{3.34}\\
    \frac{x''_3}{x''_4}=\omega = \frac{x'''_3}{x'''_4} &,&\;\;\;\;\;
    \frac{x''_{34}}{x''_4}=0=\frac{x'''_{34}}{x'''_4},
\label{3.35}\\
    \frac{x'_4}{x'_3}=\frac{1}{\omega}=\frac{x'''_{13}x'''_{24}}
                     {\omega x'''_{14}x'''_{23}}&,&\;\;\;\;\;
    \frac{x'_{34}}{x'_3}=0
    =\frac{x'''_{12}x'''_{34}}{\omega^{1/2} x'''_{14}x'''_{23}}.
\label{3.36}
\end{eqnarray}

Next we consider Eqs.~(\ref{2.14}) - (\ref{2.17}).
We rewrite these relations with $x^*_1$, $x^*_3$ and $x^*_{13}$.
Then Eqs.~(\ref{2.14}) and (\ref{2.17}) give the same relation,

\begin{eqnarray}
 \frac{x_{13}}{x_3}\frac{x'_3}{x'_{13}}
       \frac{x''_{13}}{x''_1}\frac{x'''_1}{x'''_{13}}&=&1 ,
\label{3.37}
\end{eqnarray}

\noindent
and Eq.~(\ref{2.15}) gives the relation,

\begin{eqnarray}
 \frac{x_{13}}{x_1}\frac{x'_1}{x'_{13}}
       \frac{x''_{13}}{x''_1}\frac{x'''_1}{x'''_{13}}&=&1 ,
\label{3.38}
\end{eqnarray}

\noindent
and Eq.~(\ref{2.16}) gives the relation

\begin{eqnarray}
 \frac{x_{13}}{x_3}\frac{x'_3}{x'_{13}}
       \frac{x''_{13}}{x''_3}\frac{x'''_3}{x'''_{13}}&=&1 .
\label{3.39}
\end{eqnarray}

In order for Eqs.~(\ref{3.37}) and (\ref{3.38}) to be consistent,
the condition

\begin{eqnarray}
  \frac{x_1}{x_3}=\frac{x'_1}{x'_3}
\label{3.40}
\end{eqnarray}

\noindent
 must be satisfied. This gives

\begin{eqnarray}
  \frac{ \cos^{1/N}{\theta_1} \cos^{1/N}{\theta_2}}
  {\sin^{1/N}{\theta_1} \sin^{1/N}{\theta_2} \cos^{1/N}{a_3}}
  = \frac{ \cos^{1/N}{\theta_1} \cos^{1/N}{(\pi-\theta_6)} }
   {\sin^{1/N}{\theta_1} \sin^{1/N}{(\pi-\theta_6)}
   \cos^{1/N}{a'_3}}  .
\label{3.41}
\end{eqnarray}

\noindent Taking the $N$-th power, we have

\begin{eqnarray}
   \frac{\cos{\theta_1} \cos{\theta_2} }
    {\sin{\theta_1} \sin{\theta_2} \cos{a_3} }
 =\frac{\cos{\theta_1} \cos(\pi-\theta_6) }
    {\sin{\theta_1} \sin(\pi-\theta_6) \cos{a'_3} } .
\label{3.42}
\end{eqnarray}

\noindent 
Substituting the spherical trigonometry relations

\begin{eqnarray}
 \cos{a_3}=&&\frac{\cos{\theta_3}+\cos{\theta_1} \cos{\theta_2}}
    {\sin{\theta_1} \sin{\theta_2} },  \nonumber \\
 \cos{a'_3}=&&\frac{\cos{(\pi-\theta_4)}+\cos{\theta_1}
    \cos{(\pi-\theta_6)}}
    {\sin{\theta_1} \sin{(\pi-\theta_6)} } \nonumber \\
     =&&-\frac{\cos{\theta_4}+\cos{\theta_1} \cos{\theta_6}}
    {\sin{\theta_1} \sin{\theta_6} },  \nonumber
\end{eqnarray}

\noindent
into Eq.~(\ref{3.42}), we have the constraint

\begin{eqnarray}
   \cos{\theta_2} \cos{\theta_4}=\cos{\theta_3} \cos{\theta_6} .
\label{3.43}
\end{eqnarray}

Next, in order for Eqs.~(\ref{3.37}) and (\ref{3.39}) to be consistent,
the following condition

\begin{eqnarray}
    \frac{x''_1}{x''_3}=\frac{x'''_1}{x'''_3} .
\label{3.44}
\end{eqnarray}

\noindent
must be satisfied. 

\noindent
Taking the $N$-th power, we have

\begin{eqnarray}
  \frac{ \cos(\pi-\theta_3) \cos{\theta_5} }{\sin{(\pi-\theta_3)}
  \sin{\theta_5} \cos{a''_3} }
  = \frac{ \cos{\theta_2} \cos{\theta_5}}
   {\sin{\theta_2} \sin{\theta_5} \cos{a'''_3}} .
\label{3.45}
\end{eqnarray}

\noindent 
Substituting the spherical trigonometry relations

\begin{eqnarray}
 \cos{a''_3}=&&\frac{\cos{(\pi-\theta_4)}+\cos{\theta_5}
    \cos{(\pi-\theta_3)}}
    {\sin{\theta_5} \sin{(\pi-\theta_3)} } \nonumber \\
    =&&-\frac{\cos{\theta_4}+\cos{\theta_5}\cos{\theta_3}}
    {\sin{\theta_5} \sin{\theta_3} },  \nonumber \\
 \cos{a'''_3}=&&\frac{\cos{\theta_6}+\cos{\theta_2} \cos{\theta_5}}
    {\sin{\theta_2} \sin{\theta_5} }, \nonumber
\end{eqnarray}

\noindent
into Eq.~(\ref{3.45}), we have the same constraint as Eq.~(\ref{3.43}).
Therefore, the relations given by Eqs.~(\ref{3.37}) - (\ref{3.39})
become equivalent to Eqs.~(\ref{3.38}) and (\ref{3.43}).
While, the $N$-th power of Eq.~(\ref{3.38}) is
automatically satisfied
by using Eq.~(\ref{3.15}) - (\ref{3.30}). In this way,
Eq.~(\ref{2.14}) - (\ref{2.17}) give one additional constraint,
Eq.~(\ref{3.43}).

Therefore, the parametrization of Eqs.~(\ref{3.15}) - (\ref{3.30})
and the additional constraints provied by Eqs.~(\ref{2.18})
and (\ref{3.43}) give the solution of the tetrahedron equation.

One of the non-trivial numerical examples is
$\theta_1=1.188378$, $\theta_2=1.399930$, $\theta_3=1.226731$,
$\theta_4=1.839644$, $\theta_5=0.108534$, $\theta_6=1.705095$.
$\stackrel{\frown}{AB}=1.225147$,
$\stackrel{\frown}{AC}=1.117028$,
$\stackrel{\frown}{BC}=1.088198$,
$\stackrel{\frown}{AD}=1.307240$,
$\stackrel{\frown}{AE}=1.122042$,
$\stackrel{\frown}{DE}=1.128726$,
$\stackrel{\frown}{BF}=2.323531$,
$\stackrel{\frown}{DF}=2.348327$,
$\stackrel{\frown}{BD}=0.082093$,
$\stackrel{\frown}{CF}=1.235333$,
$\stackrel{\frown}{EF}=1.219601$,
$\stackrel{\frown}{CE}=0.103396$.
We can easily check the relations
$\stackrel{\frown}{AB}+\stackrel{\frown}{BD}=\stackrel{\frown}{AD}$,
$\stackrel{\frown}{AC}+\stackrel{\frown}{CE}=\stackrel{\frown}{AE}$,
$\stackrel{\frown}{BC}+\stackrel{\frown}{CF}=\stackrel{\frown}{BF}$,
$\stackrel{\frown}{DE}+\stackrel{\frown}{EF}=\stackrel{\frown}{DF}$,
and the fundamental
spherical trigonometric relations, Eqs.(\ref{2.28}) - (\ref{2.30}), for
$\triangle ABC$, $\triangle ADE$, $\triangle BDF$ and $\triangle CEF$
in this numerical example.

\unitlength 3pt
\begin{picture}
(75,75)

\bezier{600}(25,0)(30,33)(60,68)
\bezier{600}(23,18)(62,8)(88,23)
\bezier{600}(43,68)(74,15)(80,-2)
\bezier{600}(26,42)(62,8)(90,0)

\bezier{80}(30,21)(32,20)(33,16)
\bezier{80}(34,30)(37,28)(40,30)
\bezier{80}(54,14)(55,18)(56,18)
\bezier{80}(53,59)(55,56)(53,52)
\bezier{80}(70,21)(73,21)(77,18)
\bezier{80}(80,-1)(82,0)(83,2)

\put (34,18){$\theta_1$}
\put (50,16){$\theta_2$}
\put (36,26){$\theta_3$}
\put (55,55){$\theta_4$}
\put (82,-2){$\theta_5$}
\put (74,21){$\theta_6$}

\put(26,18){$A$}
\put(31,31){$B$}
\put(58,10){$C$}
\put(45,55){$D$}
\put(67,17){$E$}
\put(78,6){$F$}

\put (29,16){\circle*{2}}
\put (36,33){\circle*{2}}
\put (50,56){\circle*{2}}
\put (60,14){\circle*{2}}
\put (72,16){\circle*{2}}
\put (77,5){\circle*{2}}

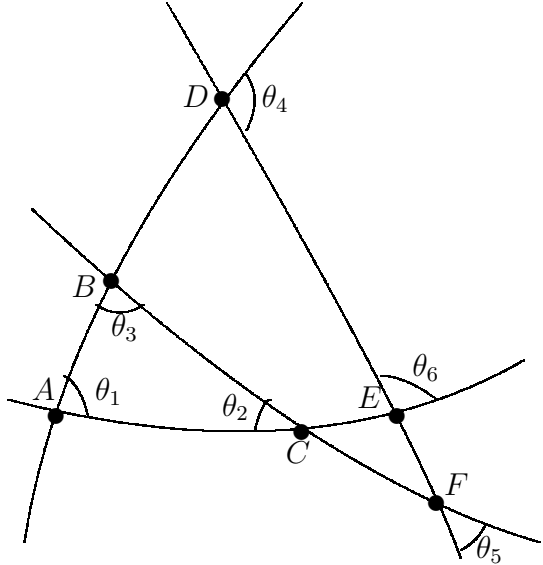
\put(20,-10){{\bf Fig.5~}~ Angles and arcs on the sphere}

\end{picture}

\vspace{10mm}


\section{Summary}

We understand the 2-dim integrable statistical model well
in the sense that we can systematically construct many integrable
statistical models and solve them to find the partition functions.
However we have only a few 3-dim integrable statistical models,
so that it will be necessary to find as many solutions
before we investigate the mathematical structure of the
3-dim integrable model.

In this paper, we first clarified the structure
of the Bazhanov-Baxter model
of the 3-dim $N$-state integrable model.
There are the two essential points,
i) the cubic symmetries, ii) the spherical trigonometry
parametrization, to understand the structure of the
Bazhanov-Baxter model.
Next we proposed two approaches to find a candidate as a
solution of the tetrahedron equation, and we found a new solution.

Our solution may be useful for understanding the mathematical
structure of the tetrahedron equation.
There may exist another new solution that can be found by considering the
higher
spin representation of the spherical trigonometry relation.

\vskip 30mm

\section*{Acknowledgements}

One of the authors (K. S. ) is grateful to the Academic Research Fund
at Tezukayama University for financial support.


\end{document}